\documentclass[pra,aps,showpacs,twocolumn,superscriptaddress]{revtex4-1}
\usepackage{amsmath}
\usepackage{amssymb}
\usepackage{graphicx}
\usepackage{epstopdf}
\usepackage{CJK}
\usepackage{easybmat}
\usepackage[bookmarks=false]{hyperref}
\hypersetup{colorlinks=true,citecolor=blue,linkcolor=blue,urlcolor=blue,pdfstartview=FitH,bookmarksopen=true}
\newcommand{\ket}[1]{|#1\rangle}

\usepackage{color,soul}
\usepackage{times,txfonts}
\setstcolor{blue}
\setulcolor{red}
\makeatletter
\usepackage{xcolor}

\newcommand{\Rmnum}[1]{\expandafter\@slowromancap\romannumeral #1@}
\makeatother
\begin{document}
\title{Single-shot realization of nonadiabatic holonomic quantum gates in decoherence-free subspaces}
\author{P. Z.  Zhao}
\affiliation{Department of Physics, Shandong University, Jinan 250100, China}
\author{G. F. Xu}
\email{Email: sduxgf@163.com}
\affiliation{Department of Physics, Shandong University, Jinan 250100, China}
\affiliation{Department of Physics and Astronomy, Uppsala University,
Box 516, Se-751 20 Uppsala, Sweden}
\author{Q. M. Ding}
\affiliation{Department of Physics, Shandong University, Jinan 250100, China}
\author{Erik Sj\"oqvist}
\email{e-mail: erik.sjoqvist@physics.uu.se}
\affiliation{Department of Physics and Astronomy, Uppsala University,
Box 516, Se-751 20 Uppsala, Sweden}
\author{D. M. Tong}
\email{Email: tdm@sdu.edu.cn}
\affiliation{Department of Physics, Shandong University, Jinan 250100, China}
\affiliation{Department of Physics, Dalian Maritime University, Dalian 116026, China}
\date{\today}

\begin{abstract}
Nonadiabatic holonomic quantum computation in decoherence-free subspaces has attracted
increasing attention recently, as it allows for high-speed implementation and combines both
the robustness of holonomic gates and the coherence stabilization of decoherence-free subspaces.
Since the first protocol of nonadiabatic holonomic quantum computation in decoherence-free
subspaces, a number of schemes for its physical implementation have been
put forward. However, all previous schemes require two noncommuting gates to realize
an arbitrary one-qubit gate, which doubles the exposure time of gates to error sources as well
as the resource expenditure. In this paper, we propose an alternative  protocol for nonadiabatic
holonomic quantum computation in decoherence-free subspaces, in which an arbitrary one-qubit
gate in decoherence-free subspaces is realized by a single-shot implementation. The present
protocol not only maintains the merits of the original protocol, but also avoids the extra work
of combining two gates to implement an arbitrary
one-qubit gate and thereby reduces the exposure time to various error sources.
\end{abstract}
\maketitle
\section{Introduction}
Quantum computers exploit the fundamental principles of coherent superposition and
quantum entanglement to provide an efficient solution to certain computational tasks,
such as factoring large integers \cite{Shor} and searching unsorted databases \cite{Grover}.
The implementation of circuit-based quantum computation relies on the ability to realize a
universal set of accurately controllable building blocks, including arbitrary one-qubit gates
and a nontrivial two-qubit gate \cite{Bremner}. However, control errors that accumulate in the
operation processing pose a serious challenge in realizing quantum computation. To tackle
this, adiabatic and nonadiabatic holonomic quantum computations have been proposed.
Holonomic gates depend only on evolution paths but not on evolution details, making them
robust against control errors.

Berry phases \cite{Berry} were first exploited to realize
quantum computation, known as adiabatic geometric quantum computation \cite{Jones}.
Such an idea was then generalized to adiabatic non-Abelian geometric phases \cite{Wilczek}, based on which adiabatic holonomic quantum computation was proposed \cite{Zanardi,Duan,Wu2005}. Adiabatic Abelian and non-Abelian geometric phases provide useful tools to quantum gates
that are robust against control errors. However, a challenge for these implementations is the
long run time needed for adiabatic evolution, which makes the gates vulnerable to
environment-induced decoherence and thereby hinders experimental realization. To resolve
this problem, nonadiabatic geometric quantum computation \cite{Wang,Zhu} based on
nonadiabatic Abelian geometric phases \cite{Aharonov} has been put forward, and later
generalized to nonadiabatic holonomic quantum computation \cite{E2012,Xu2012} based
on nonadiabatic non-Abelian geometric phases \cite{Anandan}. Since nonadiabatic holonomic
quantum computation has the merits of both short run-time and robustness against control
errors \cite{Johansson2012,Jing2017}, it has received increasing attention recently.
Up to now, nonadiabatic holonomic quantum computation has been well-developed in both theory
\cite{Spiegelberg2013,Zhang2014,Xu2014PRA,Xu2014SR,Liang2014,Zhou2015,Xue2015,Xue2016,
Pyshkin2016,Zhang2015,G2015,Xu2015,E2016,Sun2016,Liang2016,You2016,Xue2016A1,Xue2016A2}
and experiment \cite{Abdumalikov,Long,Arroyo,Duan2014}.

Apart from control errors, decoherence caused by the interaction between a quantum system
and its environment is another important challenge to realize quantum computation. To obtain
quantum gates that are robust against both control errors and decoherence, the combination
of nonadiabatic holonomic quantum computation \cite{E2012} and decoherence-free subspaces \cite{Duan1997,Zanardi1997,Lidar1998} is a promising strategy. Yet, such a combination
is nontrivial since only the decoherence-free subspaces that are compatible with the conditions
of nonadiabatic holonomic gates can be used to protect nonadiabatic holonomic quantum
computation. After a great effort, the first protocol of nonadiabatic holonomic quantum computation in
decoherence-free subspaces has been developed in Ref.~\cite{Xu2012}, and a number of
implementation schemes in various physical systems, such as trapped ions \cite{Liang2014},
nitrogen-vacancy centers \cite{Zhou2015} and superconducting circuits \cite{Xue2015,Xue2016},
have been proposed recently. Nonadiabatic holonomic quantum computation in decoherence-free
subspaces has the merits of short run-time and resilience to both control errors
and decoherence.

However, all previous schemes of nonadiabatic holonomic quantum computation in decoherence-free subspaces require two noncommuting one-qubit gates to realize an arbitrary one-qubit gate.
It doubles the exposure time of the gates to errors, as well as the resource expenditure
needed for combining two holonomic gates to achieve an arbitrary one-qubit gate. This motivates
us to consider a revised protocol to further improve the efficiency of nonadiabatic holonomic
quantum computation in decoherence-free subspaces. Noting that the
single-shot implementation approach has been used to realize nonadiabatic holonomic gates in a closed quantum system \cite{Xu2015,E2016}, we find that a similar approach can be also applied to nonadiabatic holonomic quantum computation in decoherence-free subspaces by properly choosing the system Hamiltonian.

In this paper, we propose an alternative protocol of nonadiabatic holonomic quantum computation in
decoherence-free subspaces, in which an arbitrary one-qubit gate is directly realized by a
single-shot implementation. The present protocol not only maintains the merits of the
original protocol, but also avoids the extra work of combining two gates to implement an arbitrary one-qubit gate and thereby reduces the exposure time to various error sources.

The paper is organized as follows. In Sec. \Rmnum {2}, nonadiabatic holonomic quantum
computation in decoherence-free subspaces is briefly reviewed. In Sec. \Rmnum {3}, the present
protocol is demonstrated by a physical model of a $N$-qubit system interacting with a dephasing
environment. Section \Rmnum {4} is the conclusion.

\section{Nonadiabatic holonomic quantum computation in decoherence-free subspaces}
We first recapitulate the main idea of nonadiabatic holonomic quantum computation in
decoherence-free subspaces.

Any real quantum system inevitably interacts with its environment. The dynamics of an open
quantum system cannot be described by a unitary operator due to the interaction, and the
states of the system are affected by decoherence  in general. However, if the interaction
between the quantum system and its environment possesses some symmetry, there may
exist subspaces that are immune to decoherence and the states in these decoherence-free
subspaces evolve unitarily. Then, nonadiabatic holonomic quantum computation can be
realized in the decoherence-free subspace if it contains a smaller subspace satisfying the
holonomic conditions.

To describe this idea in some more detail, we use $H$ to denote the Hamiltonian of the
quantum system under consideration, which may be time dependent, and $H_I$ to denote
the interaction between the quantum system and its environment. $H_I$ can be generally written
as $H_I=\sum_\alpha S_\alpha\otimes B_\alpha$, where $S_\alpha$ and $B_\alpha$ are
operators acting on the system and the environment, respectively. If all $S_\alpha$ have a
common set of time-independent degenerate eigenvectors $\{\ket{\psi_1}, \ket{\psi_2},
\cdots, \ket{\psi_K}\}$, which comprise an invariant subspace of $H$, denoted as
$\mathcal{S}^D$, i.e., satisfying the conditions,
\begin{align}
(a)~~&S_\alpha |\psi_{k}\rangle=\lambda_\alpha|\psi_{k}\rangle,~{\text {for all}} ~~\alpha, \notag\\
(b)~~&H|\psi_{k}\rangle\in\mathcal{S}^D,~k=1,2,\cdots, K,
\end{align}
where $\lambda_\alpha$ is a degenerate eigenvalue of $S_\alpha$, then $\mathcal{S}^D$
defines a $K$-dimensional decoherence-free subspace. In this case, a quantum state initially
prepared in the subspace $\mathcal{S}^D$ will undergo a unitary evolution with evolution
operator $U(t)=\text{T}\exp{\left(-i\int_0^tHdt\right)}$,  and remain in the subspace for the
whole evolution time. If there is a smaller subspace $\mathcal{S}^L = \textrm{Span}\{\ket{\psi_{k_1}}, \ket{\psi_{k_2}}, \cdots, \ket{\psi_{k_M}}\} \subset \textrm{Span}\{\ket{\psi_1}, \ket{\psi_2}, \cdots, \ket{\psi_K}\}$, satisfying the additional two conditions,
\begin{align}
(c)~~&U(\tau)\bigg(\sum^{M}_{i=1}|\psi_{k_i}\rangle\langle\psi_{k_i}|\bigg)U^\dagger(\tau)=
\sum^{M}_{i=1}|\psi_{k_i}\rangle\langle\psi_{k_i}|,\notag\\
(d)~~&\langle\psi_{k_i}|U(t)^\dag HU(t)|\psi_{k_j}\rangle=0,~~i,j=1,2,\dots,M,
\end{align}
where  $\tau$ is the evolution period, then nonadiabatic holonomic quantum computation
in decoherence-free subspaces can be realized by encoding computational qubits into the
$M$-dimensional subspace $\mathcal{S}^{L}$ of the $K$-dimensional decoherence-free
subspace $\mathcal{S}^{D}$ \cite{Xu2012}.

The model used in nonadiabatic holonomic quantum computation in decoherence-free
subspaces consists of $N$ physical qubits interacting collectively with a dephasing environment.
By using three neighboring physical qubits undergoing collective dephasing to encode one logical
qubit, a universal set of  holonomic gates is obtained. However, all previous schemes
\cite{Xu2012,Liang2014,Zhou2015,Xue2015,Xue2016} of
nonadiabatic holonomic quantum computation in decoherence-free subspaces require
two noncommuting gates to realize an arbitrary one-qubit gate.

\section{The protocol}
We now put forward an alternative protocol of nonadiabatic holonomic quantum computation
in decoherence-free subspaces, in which an arbitrary one-qubit gate can be obtained by a
single-shot implementation. The model used in the present protocol is $N$ physical
qubits interacting collectively with a dephasing environment \cite{Wu2005,Xu2012}. For the
dephasing environment, the interaction Hamiltonian is described by
\begin{align}
H_{I}=S_N\otimes B, \label{interaction}
\end{align}
where $S_N=\sum_{k}\sigma^{z}_{k}$ is a collective spin operator with $\sigma^{z}_{k}$
being the Pauli $z$ operator acting on the $k$th qubit, and $B$ is an arbitrary environment
operator. The Hamiltonian used in the present protocol reads
\begin{align}
H_N=\sum_{k<l}\bigg(J^{x}_{kl}R^{x}_{kl}+J^{y}_{kl}R^{y}_{kl}\bigg)+\sum_{m}J^{z}_{m}\sigma^{z}_{m}, \label{hamiltonian}
\end{align}
where $J^{x}_{kl}$ and $J^{y}_{kl}$ are real-valued controllable coupling parameters defining the
strengths of the $XY$ interaction $R^{x}_{kl} = \frac{1}{2} \left(\sigma^{x}_{k} \sigma^{x}_{l} +
\sigma^{y}_{k} \sigma^{y}_{l}\right)$ and the Dzialoshinski-Moriya interaction $R^{y}_{kl} =
\frac{1}{2}\left(\sigma^{x}_{k}\sigma^{y}_{l}-\sigma^{y}_{k}\sigma^{x}_{l}\right)$
\cite{Dzyaloshinsky,Moriya}, respectively, and $J^{z}_{m}$ is a real-valued parameter describing
the strength of a local field affecting the $m$th physical qubit. Here, $\sigma^{\nu}_m$
represents the Pauli $\nu$ operator ($\nu=x,y,z$) acting on the $m$th physical qubit.
This kind of Hamiltonian has been widely used in literature
\cite{Hartmann2007,Trif2008,Trif2010,R2012,Trifunovic2012,Mousolou2014,Mousolou2014NJP}.
Compared with the Hamiltonian used in the original protocol \cite{Xu2012}, the Hamiltonian used
here includes the local field term $\sum_{m}J^{z}_{m}\sigma^{z}_{m}$. This term plays an
important role for realizing an arbitrary one-qubit gate by a single-shot implementation.
It provides an off-resonant coupling between computational
states and logical auxiliary states, which makes the rotation
angle variable.

To realize nonadiabatic holonomic quantum computation in decoherence-free subspaces, a
universal set of quantum gates is needed. In the following, we demonstrate how to realize
an arbitrary one-qubit holonomic gate by a single-shot implementation and how to realize
an entangling two-qubit holonomic gate with the Hamiltonian expressed by Eq.~(\ref{hamiltonian}).

\subsection{One-qubit gates}
To realize an arbitrary one-qubit holonomic gate in a decoherence-free subspace by a single-shot implementation, we need to consider the quantum system of three physical qubits interacting
collectively with a common dephasing environment. In the case of $N=3$, the interaction Hamiltonian
and the system Hamiltonian are specified as $H_{I}~(=S_3\otimes B)$ and $H_3$, respectively.
For this system, there is a three-dimensional subspace,
\begin{align}
\mathcal{S}^{D}_{1}={\text {Span}}\{|100\rangle,|010\rangle,|001\rangle\},
\end{align}
which satisfies conditions (a) and (b), being a decoherence-free subspace. Here, $|0\rangle$
and $|1\rangle$ represent the eigenvectors of the Pauli $z$ operator, corresponding to eigenvalues
$+1$ and $-1$, respectively.

We choose $H_3$ to be time independent and express its nonzero parameters as
\begin{align}
J^{x}_{12}=&J\cos\phi\sin\frac{\theta}{2}\cos\varphi,\notag\\
J^{y}_{12}=&-J\cos\phi\sin\frac{\theta}{2}\sin\varphi,\notag\\
J^{x}_{13}=&-J\cos\phi\cos\frac{\theta}{2},\notag\\
J^{z}_{2}=&J^{z}_{3}=J\sin\phi, \label{condition}
\end{align}
where $J$, $\phi$, $\theta$, and $\varphi$ are time-independent parameters. Here, a key point is the choice of local fields, two of which should be the same while the third is put to zero. By inserting Eq.~(\ref{condition}) into Eq.~(\ref{hamiltonian}), and using the
Pauli operators $\sigma^{x}_{\mu}= |0\rangle_\mu\langle 1|+|1\rangle_\mu\langle 0|$,
$\sigma^{y}_{\mu}= i(|1\rangle_\mu\langle 0|-|0\rangle_\mu\langle 1|)$, and
$\sigma^{z}_{\mu}= |0\rangle_\mu\langle 0|-|1\rangle_\mu\langle 1|)$ pertaining to each
of the physical qubits, we obtain
an explicit expression of Hamiltonian $H_3$.  For simplicity, we let
$|a\rangle=|100\rangle$, $|0\rangle_{L}=|010\rangle$, and $|1\rangle_{L}=|001\rangle$.
We then have
\begin{align}
H_3 = & J\cos\phi\left(\sin\frac{\theta}{2}e^{i\varphi}|a\rangle_{L}\langle0|-\cos\frac{\theta}{2}|a\rangle_{L}\langle1|+\mathrm{H.c.}\right)
\notag\\&+2J\sin\phi|a\rangle\langle a|.
\label{hamiltonian1a}
\end{align}
It can be further rewritten as
\begin{align}
H_3=J\cos\phi(|a\rangle\langle b|+|b\rangle\langle a|)+2J\sin\phi|a\rangle\langle a|,
\label{hamiltonian1}
\end{align}
where
\begin{eqnarray}
|b\rangle=&\sin\frac{\theta}{2}e^{-i\varphi}|0\rangle_{L}-\cos\frac{\theta}{2}|1\rangle_{L}.
\label{bL1}
\end{eqnarray}

We use $|d\rangle$ to donate the dark state, i.e., the zero-energy eigenstate of $H_3$
\begin{eqnarray}
|d\rangle=&\cos\frac{\theta}{2}|0\rangle_{L}+\sin\frac{\theta}{2}e^{i\varphi}|1\rangle_{L},
\label{bL2}
\end{eqnarray}
which is orthogonal to $|a\rangle$ and $|b\rangle$.

It is easy to verify that the smaller subspace,
\begin{align}
\mathcal {S}^L_{1}={\text {Span}}\{|b\rangle,|d\rangle\}={\text {Span}}\{|0\rangle_{L},|1\rangle_{L}\},
\end{align}
satisfies conditions (c) and (d) if the evolution period $\tau_{1}$ is taken as
\begin{align}
J\tau_{1}=\pi. \label{period1}
\end{align}
Indeed, since $H_3$ is time independent, implying $U(t)^\dag H_3U(t)$$=H_3$, it is straightforward to have $\langle p|U(t)^\dag H_3U(t)|q\rangle$$=\langle p|H_3|q\rangle=0$, $p,q\in\{a,b\}$, i.e., condition (d) is satisfied. One will soon see that condition (c) is satisfied too. In this case, the subspace $\mathcal {S}^L_{1}$ can be used as the computational space, and the logic qubit is encoded in it, while $|a\rangle$ acts as an ancillary state.

The evolution operator in the decoherence-free subspace $\mathcal{S}^{D}_{1}$  can be expressed as $U_{1}(t)=\exp(-iH_{3}t)$. For $t=\tau_1$, there is $U_{1}(\tau_1)=e^{-iH_{3}\tau_1}$. By using Eq.~(\ref{period1}) and the identity $2|a\rangle
\langle a| = (|a\rangle\langle a|+|b\rangle\langle b|)+(|a\rangle\langle a|-|b\rangle\langle b|)$,
we find $U_{1}(\tau_1)=\exp\left[-i\pi\sin\phi(|a\rangle\langle a|+|b\rangle\langle b|)-i\pi A\right]$
with $A= \cos\phi (|a\rangle\langle b|+|b\rangle\langle a|)+\sin\phi(|a\rangle\langle a| -
|b\rangle\langle b|)$. We see that $[A,~ |a\rangle\langle a|+|b\rangle\langle b|]=0$, and
$A^{2n}= |a\rangle\langle a|+|b\rangle\langle b|,~A^{2n+1}=A,$ for $n=1,2,...$, which implies
\begin{align}
U_{1}(\tau_1)=&e^{-i\pi\sin\phi(|a\rangle\langle a|+|b\rangle\langle b|)}e^{-i\pi A}
\notag\\
= & e^{-i(\pi+\pi\sin\phi)}(|a\rangle\langle a|+|b\rangle\langle b|)+|d\rangle\langle d|.
\end{align}
Clearly, $U_1(\tau_1)$ maps states in the subspace $\mathcal {S}^L_{1}$ into the subspace, i.e., condition $(c)$ is indeed satisfied.

In the basis $\{|a\rangle,|b\rangle,|d\rangle\}$, the unitary operator can be written as
\begin{align}
U_{1}(\tau_{1})=\left(
  \begin{array}{ccc}
   e^{-i\gamma_{1}} & 0 & 0\\
   0 & e^{-i\gamma_{1}} & 0\\
   0 & 0 & 1\\
  \end{array}
\right),
\end{align}
where $\gamma_{1}$ is given by
\begin{align}
\gamma_{1}=\pi+\pi\sin\phi. \label{global}
\end{align}
Thus, the evolution operator projected onto the computational subspace $\mathcal {S}^L_{1}$
reads
\begin{align}
U^L_1(\tau_{1})=|d\rangle\langle d|+e^{-i\gamma_{1}}|b\rangle\langle b|. \label{u1}
\end{align}
The unitary operator $U^L_1(\tau_{1})$ represents an arbitrary one-qubit gate, for which the rotation
axis is determined by states $|b\rangle$ and $|d\rangle$, and the rotation angle is determined by
phase $\gamma_{1}$.

Therefore, to achieve a desired nonadiabatic holonomic gate in the decoherence-free subspace
$\mathcal{S}^{D}_{1}$, one first calculates $\phi$, $\theta$, and $\varphi$ by using Eqs. (\ref{bL1}), (\ref{bL2}),
and (\ref{global}), and then determines the parameters $J^{x}_{12}$, $J^{y}_{12}$, $J^{x}_{13}$,
$J^{z}_{2}$, and $J^{z}_{3}$ by using Eq. (\ref{condition}). In this way, the Hamiltonian described
by Eq.~(\ref{hamiltonian1}) is obtained, and the desired one-qubit gate can be realized by
appropriately choosing the evolution period $\tau_{1}$ satisfying Eq.~(\ref{period1}).

The above discussion shows that an arbitrary one-qubit gate can be realized by a single-shot implementation. Compared with the original protocol of nonadiabatic holonomic quantum
computation in decoherence-free subspaces as well as all the schemes of its physical implementation,
the present implementation of one-qubit gates not only maintains the merits of the previous protocol,
but also avoids the extra work
of combining two gates to implement an arbitrary one-qubit gate and thereby reduces the exposure time to various error sources.

\subsection{Entangling two-qubit gate}

To realize a universal set of nonadiabatic holonomic gates in decoherence-free subspaces, a
nontrivial two-qubit holonomic gate is needed in addition to the one-qubit
holonomic gates obtained above. Noting that our arbitrary one-qubit gates are compatible
with the nonadiabatic holonomic two-qubit gates proposed in
Refs.~\cite{Xu2012,Liang2014,Zhou2015,Xue2015,Xue2016}, we may simply combine
the present one-qubit gates with those previous two-qubit gates to form a universal set
of quantum gates. Alternatively, we can also construct an entangling two-qubit holonomic
gate by using the Hamiltonian in Eq.~(\ref{hamiltonian}), leading to a wider class of
two-qubit gates. We demonstrate how to obtain such generalized two-qubit
gates by using the Hamiltonian expressed in Eq.~(\ref{hamiltonian}). The scheme
follows closely that of the holonomic one-qubit gates discussed above.

Consider six physical qubits interacting collectively with a common dephasing environment.
In the case of $N=6$, the interaction Hamiltonian and the system Hamiltonian are
$H_{I} = S_6\otimes B$ and $H_6$, respectively. For this system, there is a six-dimensional
subspace,
\begin{align}
{{\mathcal{S}}^{D}_{2}}={\text{ Span}}\{&|010010\rangle,|010001\rangle,|001010\rangle,\notag\\
&|001001\rangle,|011000\rangle,|000011\rangle\},
\end{align}
which satisfies conditions (a) and (b), being a decoherence-free subspace.

We choose $H_6$ to be time independent and express its nonzero parameters as
\begin{align}
J^{x}_{35}=&\lambda\cos\zeta\sin\frac{\alpha}{2}\cos\beta, \notag\\
J^{y}_{35}=&-\lambda\cos\zeta\sin\frac{\alpha}{2}\sin\beta, \notag\\
J^{x}_{36}=&-\lambda\cos\zeta\cos\frac{\alpha}{2}, \notag\\
J^{z}_{5}=&J^{z}_{6}=\lambda\sin\zeta,
\label{h6parameters}
\end{align}
where $\lambda$, $\zeta$, $\alpha$, and $\beta$ are time-independent parameters. By
inserting Eq.~(\ref{h6parameters}) into Eq.~(\ref{hamiltonian}) and using the Pauli
operators $\sigma^{x}_{\mu}= |0\rangle_\mu\langle 1|+|1\rangle_\mu\langle 0|$,
$\sigma^{y}_{\mu}= i(|1\rangle_\mu\langle 0|-|0\rangle_\mu\langle 1|)$, and
$\sigma^{z}_{\mu}= |0\rangle_\mu\langle 0|-|1\rangle_\mu\langle 1|)$, we can
obtain an explicit expression for $H_6$.  We let $|00\rangle_{L}=|010010\rangle$,
$|01\rangle_{L}=|010001\rangle$, $\ket{10}_{L}=\ket{001010}$, $|11\rangle_{L} =
|001001\rangle$, $|a_{1}\rangle=|011000\rangle$, and $|a_{2}\rangle=|000011\rangle$,
where the first four states are consistent with the one-qubit code
$|0\rangle_{L}=|010\rangle$ and $|1\rangle_{L}=|001\rangle$ used before. We obtain
\begin{align}
H_{6}=&\lambda\cos\zeta\left(\sin\frac{\alpha}{2}e^{i\beta}|a_{1}\rangle_{L}\langle00|-\cos\frac{\alpha}{2}|a_{1}\rangle_{L}\langle01|
+\mathrm{H.c}.\right)\notag\\
&-\lambda\cos\zeta\left(\cos\frac{\alpha}{2}|a_{2}\rangle_{L}\langle10|-\sin\frac{\alpha}{2}e^{-i\beta}|a_{2}\rangle_{L}\langle11|
+\mathrm{H.c}.\right)\notag\\
&+2\lambda\sin\zeta(|a_1\rangle\langle a_1|-|a_2\rangle\langle a_2|).
\end{align}
It can be further rewritten as
\begin{align}
H_{6}=&\lambda\left[\cos\zeta(|a_{1}\rangle\langle b_{1}|+|b_{1}\rangle\langle a_{1}|)
+2\sin\zeta|a_1\rangle\langle a_1|\right]\notag\\
&-\lambda\left[\cos\zeta(|a_{2}\rangle\langle b_{2}|+|b_{2}\rangle\langle a_{2}|)
+2\sin\zeta|a_2\rangle\langle a_2|\right], \label{Hamiltonian2}
\end{align}
where
\begin{align}
|b_{1}\rangle = & \sin\frac{\alpha}{2}e^{-i\beta}|00\rangle_{L}-\cos\frac{\alpha}{2}
|01\rangle_{L},\notag\\
|b_{2}\rangle = & \cos\frac{\alpha}{2}|10\rangle_{L}-\sin\frac{\alpha}{2}e^{i\beta}
|11\rangle_L .
\end{align}

We use $|d_{1}\rangle$ and $|d_{2}\rangle$ to donate the dark states, i.e., the zero-energy eigenstates of $H_6$,
\begin{align}
|d_{1}\rangle = & \cos\frac{\alpha}{2}|00\rangle_{L} +
\sin\frac{\alpha}{2}e^{i\beta}|01\rangle_{L}, \notag\\
|d_{2}\rangle = & \sin\frac{\alpha}{2}e^{-i\beta}|10\rangle_{L} +
\cos\frac{\alpha}{2}|11\rangle_{L},
\end{align}
which are orthogonal to $|a_{1}\rangle$, $|a_{2}\rangle$, $|b_{1}\rangle$, and $|b_{2}\rangle$.

By following a line of argument similar to the one-qubit gates, it is easy to verify that the smaller subspace
\begin{align}
\mathcal{S}^{L}_{2}=&{\text{Span}}\{ |b_1\rangle,~|d_1\rangle,~|b_2\rangle,~|d_2\rangle\}
\notag\\
=&{\text{Span}}\{|00\rangle_{L},~|01\rangle_{L},~|10\rangle_{L},~|11\rangle_{L}\},
\notag\\ =&{\text{Span}}\{|010010\rangle,|010001\rangle,|001010\rangle,|001001\rangle\}
\end{align}
satisfies conditions (c) and (d) if the evolution period $\tau_{2}$ is taken as
\begin{align}
\lambda\tau_{2}=\pi.
\label{period2}
\end{align}
In this case, $\mathcal{S}^{L}_{2}$ can be used as the computational space, and the logical qubits are encoded in it, while $|a_{1}\rangle$ and $|a_{2}\rangle$ are utilized as two ancillary states.

The evolution operator in the decoherence-free subspace $\mathcal{S}^{D}_{2}$  can be expressed as $U_2(t)=\exp(-iH_6t)$. By further
pursuing the analogy of one-qubit gates, $U_2(\tau_2)$ can be expressed as $U_2(\tau_2)=\exp[-i\pi\sin\zeta(|a_1\rangle\langle a_1|+|b_1\rangle\langle b_1|)-i\pi A_1]\exp[i\pi\sin\zeta(|a_2\rangle\langle a_2|+|b_2\rangle\langle b_2|)+i\pi A_2]$ with $A_i= \cos\zeta(|a_i\rangle\langle b_i|+|b_i\rangle\langle a_i|)+\sin\zeta(|a_i\rangle\langle a_i|-|b_i\rangle\langle b_i|)$. Noting that $[A_i,~ |a_i\rangle\langle a_i|+|b_i\rangle\langle b_i|]=0$, and $A_i^{2n}= |a_i\rangle\langle a_i|+|b_i\rangle\langle b_i|,~A_i^{2n+1}=A_i,$ for $n=1,2,...$, we then obtain
\begin{align}
U_{2}(\tau_2)=&e^{-i\pi\sin\zeta(|a_1\rangle\langle a_1|+|b_1\rangle\langle b_1|)}e^{-i\pi A_1}e^{i\pi\sin\zeta(|a_2\rangle\langle a_2|+|b_2\rangle\langle b_2|)}e^{i\pi A_2} \notag\\
=&e^{-i(\pi+\pi\sin\zeta)}(|a_1\rangle\langle a_1|+|b_1\rangle\langle b_1|)+ e^{i(\pi+\pi\sin\zeta)}(|a_2\rangle\langle a_2|\notag\\
&+|b_2\rangle\langle b_2|)  +|d_1\rangle\langle d_1|+|d_2\rangle\langle d_2|.
\end{align}
Clearly, $U_2(\tau_2)$ maps states in the subspace $\mathcal {S}^L_{2}$ into the subspace.

In the basis $\{|a_{1}\rangle,~|a_{2}\rangle,~|b_{1}\rangle,~|d_{1}\rangle,
~|b_{2}\rangle,~|d_{2}\rangle\}$, the unitary operator takes the form
\begin{align}
U_{2}({\tau_{2}})=\left(
  \begin{array}{cccccc}
   e^{-i\gamma_{2}} & 0 & 0 & 0 & 0 & 0 \\
   0 & e^{i\gamma_{2}} & 0 & 0 & 0 &0 \\
   0 & 0 & e^{-i\gamma_{2}} & 0 & 0 & 0 \\
   0 & 0 & 0 & 1 & 0 & 0 \\
   0 & 0 & 0 & 0 & e^{i\gamma_{2}} & 0 \\
   0 & 0 & 0 &0 & 0 & 1 \\
  \end{array}
\right), \label{operator2}
\end{align}
where $\gamma_{2}$ is given by
\begin{align}
\gamma_{2}=\pi+\pi\sin\zeta.
\end{align}
The evolution operator projected onto the computational subspace $\{|b_{1}\rangle,|d_{1}\rangle,|b_{2}\rangle,|d_{2}\rangle\}$ reads
\begin{align}
U^L_2(\tau_{2})=&|d_1\rangle\langle d_1|+e^{-i\gamma_{2}}|b_1\rangle\langle b_1|\notag\\
&+|d_2\rangle\langle d_2|+e^{i\gamma_{2}}|b_2\rangle\langle b_2|. \label{u2}
\end{align}

$U^L_2(\tau_{2})$ acts as an entangling two-qubit gate in the computational subspace
$\mathcal{S}^{L}_{2}$. To see this explicitly, we note that
$U^L_2(\tau_{2}) = |0\rangle_L\langle0|\otimes\left[\exp(-i\gamma_{2}/2) \exp(i\gamma_{2}\boldsymbol{n\cdot\sigma}/2)\right]
$$+ |1\rangle_L\langle1| \otimes \left[ \exp(i\gamma_{2}/2) \exp(-i\gamma_{2}
\boldsymbol{m\cdot\sigma}/2) \right]$ with $\boldsymbol{n}=(\sin\alpha\cos\beta,$ $~\sin\alpha\sin\beta,$$~\cos\alpha)$ and $\boldsymbol{m}=(\sin\alpha\cos\beta,~\sin\alpha\sin\beta,$$~-\cos\alpha)$. Here, $\boldsymbol{\sigma}$$=(\sigma^x,\sigma^y,\sigma^z)$ are Pauli operators, acting on $|0\rangle_{L}$ and $|1\rangle_{L}$.  Since $\gamma_{2}$ in
Eq.~(\ref{u2}) can take any desired value by properly choosing $\zeta$, instead of being fixed to $\gamma_{2}=\pi$ in all the previous
schemes \cite{Xu2012,Liang2014,Zhou2015,Xue2015,Xue2016}, $U_2^L (\tau_2)$ covers a wider class of holonomic two-qubit gates in the decoherence-free subspace $\mathcal{S}_2^D$.

The arbitrary one-qubit gate $U^L_1(\tau_{1})$ given in Eq. (\ref{u1}) and the entangling two-qubit gate $U^L_2(\tau_{2})$ given in Eq. (\ref{u2}) comprise a universal set of nonadiabatic holonomic gates in decoherence-free subspaces.

\section{Conclusion}

In conclusion, we have proposed an alternative protocol of nonadiabatic holonomic quantum
computation in decoherence-free subspaces, in which an arbitrary one-qubit gate is directly
realized by a single-shot implementation. The present protocol not only maintains the merits
of the original protocol, but also avoids the extra work of combining two gates to implement an arbitrary one-qubit gate and thereby reduces the exposure time to various error sources. We hope that the present protocol will be useful to find new experimentally
feasible settings that combine the ideas of holonomic quantum and decoherence-free subspaces
for robust quantum computation.

\begin{acknowledgments}
P.Z.Z. acknowledges support from the National Natural Science Foundation of China though
Grant No. 11575101. G.F.X. acknowledges support from the National Natural Science Foundation
of China through Grant No. 11605104, the Future Project for Young Scholars of Shandong
University through Grant No. 2016WLJH21, and the Carl Tryggers Stiftelse through
Grant No. 14:441. E.S. acknowledges financial support from the Swedish Research Council (VR)
through Grant No. D0413201. D.M.T. acknowledges support from the National Basic Research
Program of China through Grant No. 2015CB921004.
\end{acknowledgments}

\end{document}